\documentclass[
aip,
jmp,
amsmath,amssymb,
reprint,
]{revtex4-1}

\usepackage{graphicx}
\usepackage{dcolumn}
\usepackage{bm}
\usepackage{subcaption}

\begin{document}

    \title[]{Elastic and Viscous Effects in Viscoelastic Flows: Elucidating the Distinct Roles of the Deborah and Weissenberg Numbers}
    
	\author{Luis G. Sarasua}
	\email{gustavo.sarasua@fcien.edu.uy}
	\affiliation{Instituto de F\'isica, Universidad de la Rep\'ublica, Uruguay.}
	
	\author{Daniel Freire Caporale}
	\affiliation{Instituto de F\'isica, Universidad de la Rep\'ublica, Uruguay.}
	
	\author{Arturo C. Marti}
	\affiliation{Instituto de F\'isica, Universidad de la Rep\'ublica, Uruguay.}
	
	\date{\today}
	
	\begin{abstract}

The interpretation of the parameters appearing in constitutive models for viscoelastic fluids is essential for
analyzing theoretical predictions and understanding the origin of phenomena observed in experiments. In this
work, we examine the physical significance of the Deborah ($De$) and Weissenberg ($Wi$) numbers, along with
other key parameters commonly used in these models. The central objective is to clarify the extent to which these dimensionless groups effectively characterise the competition between elastic and viscous effects in complex flows. While these parameters are ubiquitous in theoretical and experimental research, their interpretation is often context-dependent and prone to ambiguity. To address this, we analyse two representative scenarios: an analytical solution for unsteady planar flow and a numerical simulation of viscoelastic flow between rotating coaxial cylinders, governed by the Oldroyd-B constitutive equations. Our findings elucidate the distinct roles of these dimensionless numbers, offering guidelines for their rigorous interpretation in both analytical and numerical studies.
\end{abstract}
	
\pacs{47.50.-d, 47.57.Ng, 83.60.Bc}
        
\keywords{Viscoelastic fluids, Deborah number, Weissenberg number, Constitutive equations, Oldroyd-B model, Taylor-Couette flow}
        
\maketitle
	
\section{Introduction}
	\label{sec:intro}
	Understanding the physical meaning of the parameters appearing in constitutive 
	models for viscoelastic fluids is essential for interpreting theoretical 
	predictions and explaining experimental observations \cite{Larson,HP}. Unlike Newtonian fluids, 
	whose behaviour is well captured by the Navier--Stokes equations, viscoelastic 
	fluids exhibit more complex dynamics due to their ability to store and release 
	elastic energy. As a result, their mathematical description introduces 
	additional dimensionless numbers beyond the classical Reynolds number ($Re$). 
	Among these, the Deborah ($De$) and Weissenberg ($Wi$) numbers play a central 
	role in characterising the relative importance of elastic versus viscous 
	effects. While other non-dimensional parameters may be relevant depending on the
	complexity of the constitutive model, $De$ and $Wi$ are particularly 
	significant for our analysis. 
	
	The precise definition of the Deborah and Weissenberg numbers 
	presents significant conceptual and practical challenges. Although both are 
	fundamental to the description of viscoelastic fluid behaviour, their 
	formulations vary considerably depending on the physical context and the 
	theoretical model employed. As a result, various definitions can be found across
	the literature \cite{HP, Delay, Larson,Larson2, Poole, Huilgol, Souza}, leading to 
	ambiguities in their interpretation and application. These inconsistencies 
	reflect the inherent complexity of viscoelastic fluids, where the interplay 
	between elastic and viscous effects strongly depends on the flow history, 
	geometry, and the timescale of observation.
	
	Generally, the Deborah number is defined as the ratio between the fluid's 
	relaxation time $\lambda$ and a characteristic timescale for changes in the flow
	conditions, $t_{sc}$, in a Lagrangian sense: $De = \lambda / t_{sc}$.
    This number indicates the degree to which a material behaves elastically under a 
	given deformation rate.

    The Weissenberg number quantifies the relative importance of elastic to viscous forces. It is defined \cite{HP} as the ratio of the first normal stress difference
    to the shear stress.
     This metric reflects the local elastic \textit{response} and is flow-dependent.
    Alternatively, a kinematic definition of the Weissenberg number is frequently encountered in the literature, 
     given by $W=\lambda \dot{\gamma}_{c}$, where $\dot{\gamma}_{c}$ represents a characteristic shear rate serving as an independent control parameter.
    
    In this work, we focus primarily on the Oldroyd-B model, as it captures the essential competition between elastic and viscous effects in a simple yet robust framework. Nevertheless, the main conclusions are applicable
	to other constitutive models, including those that account for shear-thinning 
	behaviour and finite polymer extensibility. We examine the effectiveness of different dimensionless parameters 
	in capturing the relative importance of elastic effects in viscoelastic flows. 
	We focus on the Deborah and Weissenberg numbers, along with additional 
	parameters derived from a systematic non-dimensionalisation of the Oldroyd-B 
	constitutive equations. These analytical insights are contrasted with both an 
	exact solution for an unsteady planar flow and numerical simulations of the 
	transient flow between coaxial cylinders.
	
	The structure of the paper is as follows. Section \ref{Seceq} discusses the physical 
	significance and various definitions of the Deborah and Weissenberg numbers. In 
	Section \ref{SecND}, we present the Oldroyd-B constitutive model and derive a 
	dimensionless parameter that quantifies elastic effects based on the 
	non-dimensional form of the governing equations. Section \ref{SecTransient} is devoted to 
	comparing different parameters with the observed elastic responses in both 
	planar and cylindrical geometries. In Section \ref{SecOtherModels}, we assess the relevance and 
	generalizability of the findings to more complex viscoelastic models, including 
	the FENE-P, Giesekus, and Phan-Thien-Tanner models. 
	Finally, Section \ref{SecConclusions} summarises the main conclusions and outlines possible 
	directions for future work, including the extension of this analysis to more 
	realistic flow conditions.
	
	\section{Governing equations and the definitions of the Deborah and Weissenberg 
		numbers}\label{Seceq}
	
	The flow of an incompressible fluid is governed by the momentum conservation
	\begin{equation}
		\rho \frac{D {\bf v}}{Dt}=  -\nabla p + \nabla \cdot {\bf \tau}+{\bf g},
		\label{em}
	\end{equation}
	and the continuity equation 
	\begin{equation}
		\nabla \cdot {\bf v}=0.
	\end{equation}
	where $\rho$ is the (constant) density,  $p$ the pressure, $\tau$ is the extra 
	stress tensor and ${\bf g}$ is the body force due to an external field.
	When the fluid is Newtonian, the extra stress is given by \(\tau_{ij} = \tau_{ij}^{\mathsf v} \equiv  2\mu\left(E_{ij} - \tfrac{1}{3}(\nabla\cdot\mathbf{v})\delta_{ij}\right)\), where \(\mu\) is the dynamic viscosity and \(\displaystyle E_{ij}=\tfrac{1}{2}\left(\frac{\partial v_i}{\partial x_j} + \frac{\partial v_j}{\partial x_i}\right)\) is the rate-of-deformation tensor. In incompressible flows (\(\nabla\cdot\mathbf{v}=0\)) this reduces to \(\tau_{ij}^{\mathsf v}= 2\mu E_{ij}\).
    In this case Eq.~(\ref{em}) constitutes the Navier--Stokes equation. When the fluid is 
	viscoelastic, the extra stress must include an additional term $\tau^{\mathsf p}$ to account for polymer elasticity, so that $\tau = \tau^{\mathsf v}+ \tau^{\mathsf p}$.

\begin{table}[t]
\centering
\begin{tabular}{lll}
\hline
Symbol & Description & Units \\
\hline
$\mathbf{v}$ & Velocity vector & m s$^{-1}$ \\
$p$ & Pressure & Pa \\
$\tau$, $\tau^{v}$, $\tau^{p}$ & Total, viscous, and polymeric extra stresses & Pa \\
$\sigma$ & Conformation tensor & -- \\
$\lambda$ & Polymer relaxation time & s \\
$G$ & Polymer modulus & Pa \\
$\mu_s$, $\mu_p$ & Solvent and polymeric viscosities & Pa s \\
$De$ & Deborah number ($\lambda/t_{sc}$) & -- \\
$Wi$ & Stress-based Weissenberg number & -- \\
$W$ & Kinematic Weissenberg number ($\lambda\dot{\gamma}$) & -- \\
\hline
\end{tabular}
\caption{List of principal symbols used in the text.}
\label{tab:symbols}
\end{table}

\subsection{Oldroyd-B model and other constitutive equations}

As mentioned above, several constitutive models have been proposed to describe 
viscoelastic flows. The simplest model that includes both polymer elasticity and solvent viscosity is the Oldroyd-B model.
It provides a minimal
yet physically insightful framework: the total stress
is represented as the superposition of the Newtonian viscous
contribution from the solvent and the linear elastic
response of the polymer chains, modelled as Hookean
dumbbells. This structure enables the model to capture
the essential qualitative features of many viscoelastic
flows while retaining a mathematical formulation that
remains tractable for both analytical derivations and numerical
simulations.
Provided that $\tau^{\mathsf p}=G{\bf \sigma}$, its governing equations are:
	\begin{equation}
		\rho \frac{D {\bf v}}{Dt}=  -\nabla p + \mu_s \nabla^2 {\bf v} +G \nabla 
		\cdot {\bf \sigma},
		\label{b1}
	\end{equation}
	and $\nabla \cdot {\bf v}= 0$, where ${\bf \sigma}$ is the conformation tensor, $\mu_s$ is the solvent 
	viscosity, $G$ is a polymer modulus and $\rho$ is the fluid density. The conformation tensor evolves according to
	\begin{equation}
        \lambda \stackrel{\nabla}{\bf \sigma} + ( {\bf \sigma} - I) = 0,
		\label{ob3}
	\end{equation}
	where $\stackrel{\nabla}{\mathbf{\sigma}}$ is the upper-convected derivative of the conformation tensor, i.e., $\stackrel{\nabla}{\mathbf{\sigma}} \equiv \partial_t\mathbf{\sigma}+ \mathbf{v}\cdot\nabla\mathbf{\sigma}-\mathbf{\sigma}\cdot\nabla\mathbf{v} - (\nabla\mathbf{v})^T\cdot\mathbf{\sigma}$, $\lambda$ is the relaxation time and $I$ is the identity tensor. Equivalently, one can define the polymeric stress ${\bf s}=G(\sigma - I)$, with $G=\mu_p/\lambda$ and total viscosity $\mu=\mu_s+\mu_p$, where $\mu_p$ is termed polymeric viscosity.
	
	Although we focus on Oldroyd-B, similar considerations apply to more complex models. For example, the FENE-P model accounts for finite polymer extensibility. Its equations (derived by Bird et al.~\cite{Bird2}) are identical to (\ref{b1})--(\ref{ob3}) except for an extra factor $f$ in the polymer stress:
	\begin{equation}
		\rho \frac{D {\bf v}}{Dt}=  -\nabla p + \mu_s \nabla^2 {\bf v} +f G \nabla 
		\cdot {\bf \sigma},
	\end{equation}
	\begin{equation}
		\stackrel{\nabla}{\mathbf{\sigma}} = 
		\frac{f}{\lambda} ({\bf \sigma} -I),
		\label{fene3}
	\end{equation}
	and $\nabla \cdot {\bf v}= 0$, where 
	\[
	f= \frac{l^2}{l^2-\mathrm{Tr}(\sigma)},
	\] 
	with $l$ the finite extensibility parameter. 
    
    An alternative framework is the Giesekus model, which accounts for shear-thinning behavior. In this formulation, the polymeric contribution to the extra stress is denoted by ${\bf s}$, and the governing equations are given by:
	\begin{equation}
		\rho \frac{D {\bf v}}{Dt}=  -\nabla p + \mu_s \nabla^2 {\bf v} + \nabla 
		\cdot {\bf s},
		\label{Giesekus21}
	\end{equation}
	\begin{equation}
		{\bf s}+ \lambda \stackrel{\nabla}{\bf s} + \frac{\alpha \lambda}{\mu_p} {\bf s}^2=  2 \mu_p {\bf E},
		\label{Giesekus2}
	\end{equation}
	and $\nabla \cdot {\bf v=} 0$, where $\alpha$ is a rheological parameter.
	
	Lastly, the linear Phan-Thien-Tanner (PTT) model is given by the momentum Eq.~(\ref{Giesekus21}) with
	\begin{equation}
		\lambda \stackrel{\nabla}{\bf s}+ {\bf s} \Big(1+ \frac{\epsilon \lambda}{\mu_p} 
		\mathrm{Tr}({\bf s})\Big)  =  2 \mu_p {\bf E},
		\label{PTTmodel}
	\end{equation}
	where $\epsilon$ is an extensibility parameter.

	\subsection{The Deborah number}

As discussed above, the response of an Oldroyd-B fluid is governed by two independent material parameters: the relaxation time $\lambda$ and the elastic modulus $G$ (or, equivalently, the polymer viscosity $\mu_p = G\lambda$). The Deborah number is defined as $De=\lambda/t_{sc}$, where $t_{sc}$ denotes a characteristic time scale of the process in a Lagrangian sense \cite{Poole}. If $L$ represents the characteristic distance over which the flow evolves from this perspective, the time scale can be estimated as $t_{sc}=L/U$, leading to $De=\lambda U/L$.

Consider now the limit $G \rightarrow 0$ while keeping $\lambda$ and the boundary conditions fixed. In this limit the elastic forces become negligible because the last term in Eq.~(\ref{ob3}) vanishes, even though $De$ remains finite. This observation shows that $De$ does not provide an adequate measure of the contribution of elastic forces, since a proper metric for their influence should vanish as $G \to 0$, that is, when elastic effects vanish.

Consequently, any parameter intended to characterize the viscoelastic nature of an Oldroyd-B fluid must necessarily depend on both $\lambda$ and $G$.

\subsection{The Weissenberg number}\label{sec:Wi}

As noted in the introduction, the Weissenberg number can be defined as the ratio between the first normal stress difference and the shear stress \cite{HP},
\begin{equation}
    Wi=\frac{\tau_{xx}-\tau_{yy}}{\tau_{xy}} .
    \label{dwi}
\end{equation}
In many studies $De$ and $Wi$ have been treated as equivalent \cite{Poole}. Poole examined this relation using the upper-convected Maxwell (UCM) model \cite{HP} and, assuming steady shear flow, concluded that the Deborah and Weissenberg numbers are effectively interchangeable in a variety of flows.

Here we reconsider this argument using the Oldroyd-B model. For stationary shear flow between parallel plates, the first normal stress difference is $N_1=\tau_{xx}-\tau_{yy}=2G\lambda^2\dot{\gamma}^2$ and the shear stress is $\tau_{xy}=(\mu_s+\mu_p)\dot{\gamma}$ \cite{Shaqfeh}, yielding
\begin{equation}
Wi=\frac{2G\lambda^2\dot{\gamma}}{\mu_s+G\lambda}.
\label{w2}
\end{equation}
Unlike $De$, $Wi$ depends explicitly on $G$ and vanishes in the limit $G\to0$, as expected for a parameter measuring the strength of elastic effects. Since $De$ lacks this dependence, the two quantities cannot, in general, be considered equivalent.

In theoretical and numerical studies a kinematic Weissenberg number is also often introduced,
\begin{equation}
    W=\lambda\dot{\gamma}_c,
    \label{dwi_kinematic}
\end{equation}
where $\dot{\gamma}_c$ is a characteristic shear rate (we denote this definition by $W$ to distinguish it from Eq.~(\ref{dwi})). When the characteristic time scale satisfies $t_{sc}\sim1/\dot{\gamma}_c$, $W$ and $De$ are formally equivalent \cite{Poole}. However, this definition depends only on $\lambda$ and not on $G$, and therefore shares the same limitation: it remains finite as $G\to0$ and cannot by itself quantify the magnitude of elastic forces. A relation similar to Eq.~(\ref{w2}) was obtained in Ref.~\cite{Thompson} using a different approach.

\subsection{Microscopic origin of the Oldroyd-B model}
	
The kinetic theory derivation offers an alternative perspective, grounded in a microscopic physical basis, for the independence of $\lambda$ and $G$~\cite{Larson2}:
\[
\lambda = \frac{3 \pi a \mu_s}{4 k_B T \beta_L^2}, 
\quad
\mu_p = \frac{3 \pi m a \mu_s}{4 \beta_L^2},
\]
where $k_B$ is the Boltzmann constant, $T$ is the absolute temperature and $m$ is polymer concentration, and $a$ and $\beta_L^{-1}$  are characteristic lengths of the polymeric chain model~\cite{Larson2}.
  From this, the elastic modulus is $G=\mu_p/\lambda = m k_B T$.
 This result is crucial: $G$ (and $\mu_p$) are directly proportional to the polymer concentration $m$, whereas $\lambda$ is not.
    Therefore, the kinematic dimensionless numbers, $De=\lambda/t_{sc}$ and $W=\lambda \dot{\gamma}_c$, are both independent of polymer concentration. This reinforces our central argument: $De$ and $W$ alone are insufficient, as they cannot distinguish between an extremely dilute solution ($m \to 0$, $G \to 0$) and a concentrated one ($m > 0$, $G > 0$) if $\lambda$ and the flow kinematics remain the same. 
    
    From this microscopic viewpoint, $De$ and $W$ characterize the temporal response of an individual polymer chain, whereas $G$-dependant measures, as $Wi$ from Eq.~(\ref{dwi}), capture the collective elasticity of the material.

	\section{Non-dimensionalization of the governing equations}\label{SecND}
    
To derive dimensionless groups, we nondimensionalize Eqs.~(\ref{b1})-(\ref{ob3}) using a velocity scale $U$ and a length scale $\ell$, such that $\nabla \sim 1/\ell$. Based on these, a characteristic time is defined as $t_f=\ell/U$. By introducing the definitions ${\bf v'}={\bf v}/U$, ${\bf r'}={\bf r}/\ell$, $t'=t/t_f$, $p'=p/(\rho U^2)$, $Re=U\ell\rho/\mu$, $\beta=\mu_s/\mu$, and $\zeta=G/(\rho U^2)$, the governing equations are transformed into
	\begin{equation}
		\frac{D {\bf v'}}{Dt'}=  -\nabla' p' + \frac{\beta}{Re} \nabla'^2 {\bf v'} 
		+\zeta\, \nabla' \cdot {\bf \sigma},
		\label{ob1a}
	\end{equation}
	\begin{equation}
		De \Big( \frac{\partial {\bf \sigma}}{\partial t'}  + {\bf v'} \cdot \nabla' 
		{\bf \sigma} -  {\bf \sigma} \cdot \nabla'{\bf v'}  - \nabla' {\bf v'}^T  
		\cdot {\bf \sigma}\Big)  =  I-{\bf \sigma}.
		\label{ob3a}
	\end{equation}

A measure of polymer elasticity relative to solvent viscosity can be obtained by comparing the third term to the second term on the right-hand side of Eq.~(\ref{ob1a}). Since the scaling ensures that $\nabla'^2{\bf v'}$ and $\nabla' \cdot {\bf \sigma}$ are of order $O(1)$ \cite{Kundu}, this ratio is given by

	\begin{equation*}
		\Gamma = \frac{\zeta}{\beta / Re} = \frac{G\ell}{\mu_s U}.
		\label{eq:Gamma_adim}
	\end{equation*}
	
It is worth noting that the length scale $\ell$ is not determined
in a simple way by the boundary conditions. For instance, in the canonical flow past a cylinder, the relevant length scale is the boundary layer thickness rather than the cylinder radius~\cite{Kundu}.
This introduces an ambiguity in the evaluation of $\Gamma$.
To resolve this issue, we combine $\Gamma$ with the Weissenberg number $Wi$.
As discussed in Sec.~\ref{sec:Wi}, Eq.~(\ref{dwi}) provides the Weissenberg number for shear flow between parallel plates. In this geometry, the shear rate scales as $\dot{\gamma} = \partial v_x/\partial \sim U/\ell$. Consequently, the Weissenberg number can be expressed as

    \begin{equation*}
		Wi=\frac{2G\lambda^2}{\mu_s+G\lambda} \frac{U}{\ell}.\label{eq:WiAdimParallelPlates}
	\end{equation*}
	
Since both $\Gamma$ and $Wi$ compare the elastic to viscous
forces, but they depend on opposite forms on $\ell$, their product  is
independent on it. Thus, we define the parameter
	\[
	\vartheta_e = \sqrt{\frac{Wi\,\Gamma}{2}} 
	= \frac{G \lambda}{\sqrt{\mu_s(\mu_s+G\lambda)}}.
	\]
	Importantly, $\vartheta_e$ depends only on $\lambda$, $G$, and $\mu_s$, and is independent of $U$ and $\ell$. Consequently, it constitutes an intrinsic fluid property that characterises the fluid's tendency for elastic response, regardless of the flow conditions. We will show that in transient flows the overshoot is mainly determined by $\vartheta_e$ when $U$ and $\ell$ are fixed. In the following section we examine the effects of $Wi$, $\vartheta_e$, and $W$ on the viscoelastic flow.
	
	\section{Transient flows of an Oldroyd-B fluid}\label{SecTransient}
	To explore the connection between the parameters $W$, $Wi$, $\Gamma$, and $\vartheta_e$ and elastic effects, we consider start-up flows in two configurations: parallel plates and concentric cylinders. These unsteady flows exhibit a characteristic velocity overshoot due to elastic relaxation \cite{Ren}. The magnitude of this overshoot can serve as a measure of elastic strength.
	
	\begin{figure}[htb!]
		\centering
		\includegraphics[width=1.0\linewidth,keepaspectratio]{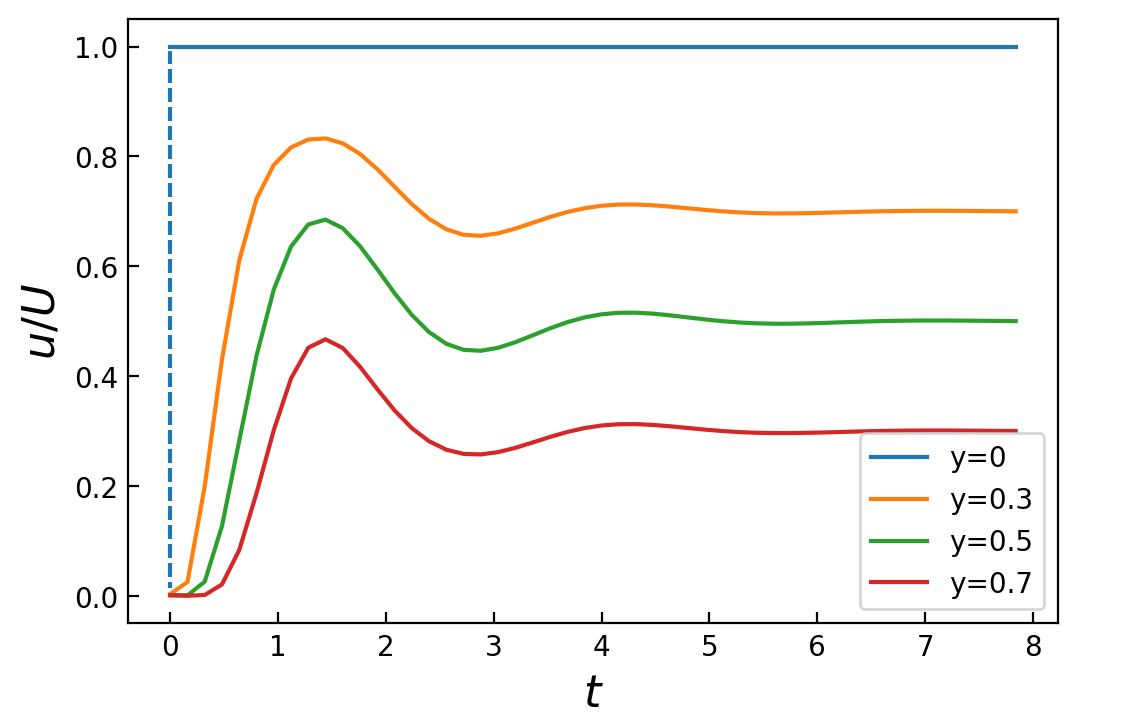}
		\caption{Dimensionless velocity component $u/U$ for different values of the vertical coordinate $y$.  Note that the velocity profile exhibits a non-monotonic behavior across the gap. 
It should also be emphasized that the flow satisfies the imposed boundary conditions.}
		\label{fig1}
	\end{figure}
	
	\begin{figure}[htb!]
		\centering
		\begin{subfigure}[b]{1.0\columnwidth}
			\centering
			\includegraphics[width=\linewidth]{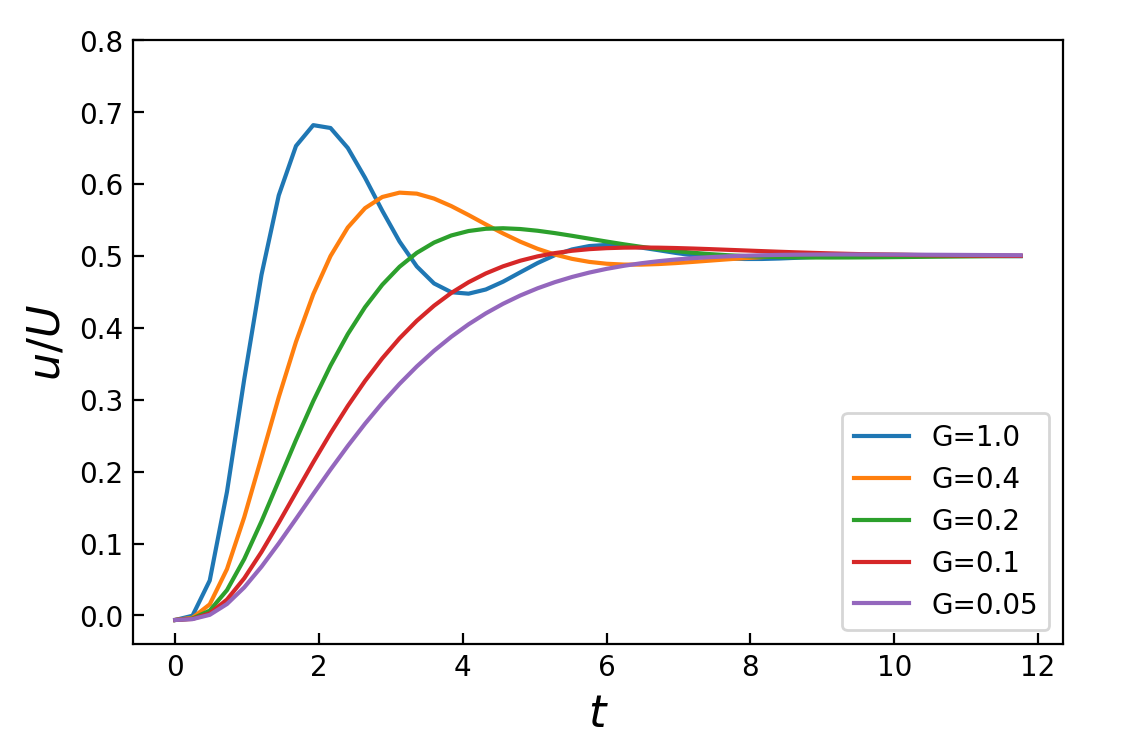}
			\caption{}
			\label{fig2}
		\end{subfigure}
		\begin{subfigure}[b]{1.0\columnwidth}
			\centering
			\includegraphics[width=\linewidth]{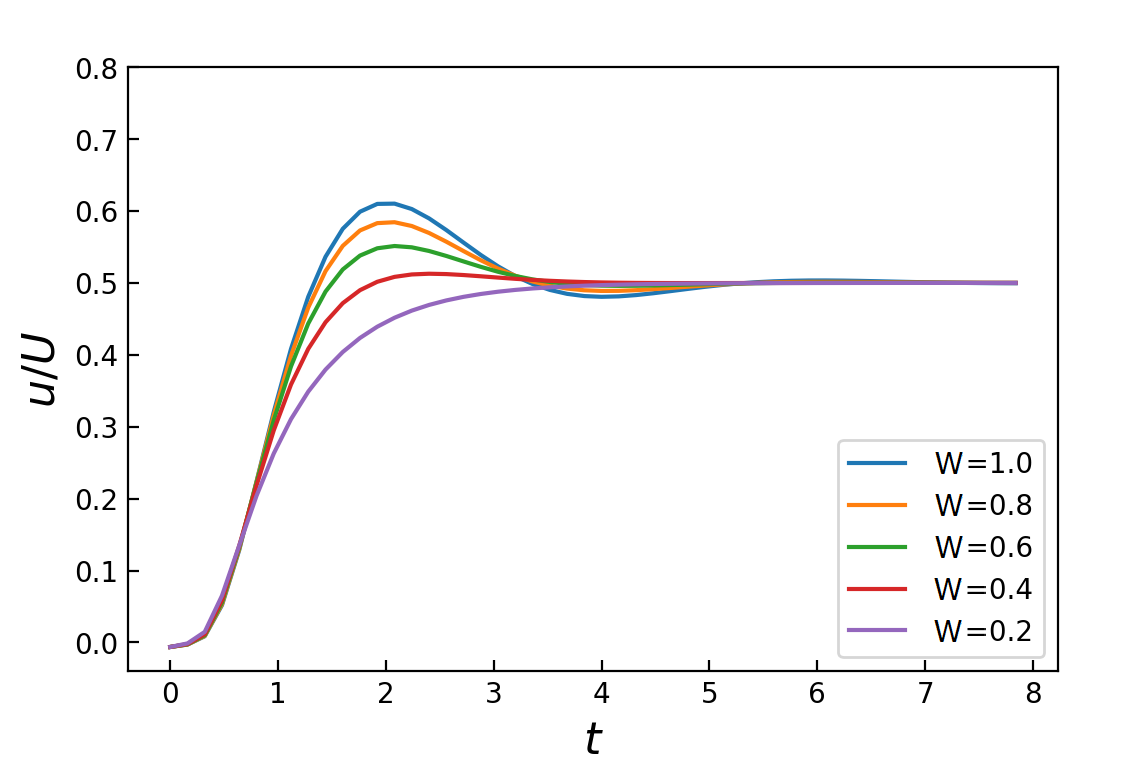}
			\caption{}
			\label{fig3}
		\end{subfigure}
		\caption{Variation of the dimensionless velocity component $u/U$ for different parameter values in the parallel-plate flow. In 
		(a), $W=\lambda U/d=2$, $\mu_s=0.3$, $\rho=4$, and $G$ takes the values shown. 
		In (b), $G=1$ and $W$ takes the values shown.}
		\label{fig23}
	\end{figure}
	
	\subsection{Flow between parallel plates}
	
	Analytical solutions for start-up Couette flow of Oldroyd-B fluids have been given by various authors \cite{Tanner,Raj,Hayat,Hayat2}. In these solutions, the fluid is between two parallel plates at $y=0$ and $y=d$, and the velocity is unidirectional, ${\bf v}=u(y,t){\bf i}$, with $\sigma=\sigma(y,t)$.
	
We consider the case in which the fluid is limited by
two plates located at $y = 0$ and $y = d$. Initially, the fluid
and the plates are at rest, and at $t = 0$ the plate at $y = 0$
suddenly starts to move at constant velocity $U$.
A common approach to obtain the solution is to eliminate
the stress tensor in order to derive an equation involving
only the velocity component $u(y, t)$. Under the
assumptions stated above, the equations of the model,
expressed in terms of the polymeric stress tensor, become
	\begin{equation}
		\rho \frac{\partial u}{\partial t}=-\frac{\partial p}{\partial x}+\mu_s 
		\frac{\partial^2 u}{\partial y^2}  +G \frac{\partial \sigma_{xy}}{\partial y} ,
		\label{ecred1}
	\end{equation}
	\begin{equation}
		\sigma_{xy}+\lambda \Big( \frac{\partial \sigma_{xy}}{\partial t} -\sigma_{yy} \frac{\partial 
			u}{\partial y} \Big)=   \frac{\mu_p}{G} \frac{\partial u}{\partial y}  ,
		\label{ecred2}
	\end{equation}
	\begin{equation}
		\sigma_{yy}+\lambda \frac{\partial \sigma_{yy}}{\partial t}  =0,
		\label{ecred3}
	\end{equation}
	\begin{equation}
		\sigma_{xx}+\lambda
        \left(
        \frac{\partial \sigma_{xx}}{\partial t}
        -2 \sigma_{xy} \frac{\partial u}{\partial y}
        \right) = 0.
	\end{equation}
	From (\ref{ecred3}) and the initial condition
    $\sigma(y,0)={\bf 0}$ 
    it follows that $\sigma_{yy}(y,t)=0$. Differentiating (\ref{ecred2}) with respect to $y$ and combining with (\ref{ecred1}) under zero pressure gradient gives:
	\begin{equation}
		\rho \frac{\partial u}{\partial t} + \rho \lambda \frac{\partial^2 u}{\partial t^2}
		-\mu \frac{\partial^2 u}{\partial y^2} -\lambda \mu_s \frac{\partial^3 u}{\partial t \partial y^2}  =0 .
		\label{ecu}
	\end{equation}
	
	An exact solution for this problem is given by Hayat {\it et al.} \cite{Hayat}.
    For convenience, we derive it here in a different way
    .
    The boundary conditions are:
	\begin{equation}
		u(y,0)= 0,\quad \frac{\partial u}{\partial t}(y,0)=0,
		\label{cb1}
	\end{equation}
	\begin{equation}
		u=U , \mbox{ at } y=0,\label{cb2}
	\end{equation}
	\begin{equation}
		u=0 , \mbox{ at } y=d.\label{cb3}
	\end{equation}
	We seek a solution of the form
	\begin{equation}
		u(y,t)=U\Big(1-\frac{y}{d}\Big) -\frac{2U}{\pi} \sum_{n=1}^\infty \frac{1}{n} 
		\sin\Big(\frac{n\pi y}{d}\Big)\, A_n(t).
		\label{ut}
	\end{equation}
	This satisfies the boundary conditions (\ref{cb2})-(\ref{cb3}) provided $A_n(0)=1$. Substituting (\ref{ut}) into (\ref{ecu}) yields the ODE for $A_n(t)$:
	\[
	\lambda \rho A_n''(t)+ \big(\rho +\lambda \mu_s n^2 \pi^2\big) A_n'(t)+  \mu n^2 \pi^2 
	A_n(t)=0.
	\]
	The solution is
	\[
	A_n(t)=\Re\{A_{n1} e^{\kappa_{n1} t}+A_{n2} e^{\kappa_{n2} t}\},
	\]
	where $\Re$ denotes the real part, and
	\[
	\kappa_{n,i}=\frac{-(\rho +\lambda \mu_s n^2 \pi^2)\pm \sqrt{(\rho +\lambda \mu_s n^2 \pi^2)^2-4\lambda \rho \mu n^2 \pi^2}}{2 \lambda \rho}.
	\]
	Imposing $A_n(0)=1$ and $A_n'(0)=0$, due to the boundary conditions from Eq.~(\ref{cb1}), gives
	\[
	A_{n1}=\frac{\kappa_{n2}}{\kappa_{n2}-\kappa_{n1}} , \quad A_{n2}=\frac{- \kappa_{n1}}{\kappa_{n2}-\kappa_{n1}}.
	\]
	Thus the exact solution is

\begin{eqnarray}
   u(y,t) &=& U\Big(1-\frac{y}{d}\Big) \nonumber \\
     &-& \frac{2U}{\pi} \sum_{n=1}^\infty \frac{1}{n} 
		\Re\Big(\frac{\kappa_{n2} e^{\kappa_{n1} t}-\kappa_{n1} e^{\kappa_{n2} t}}{\kappa_{n2}-\kappa_{n1}}\Big) \sin\Big(\frac{n\pi y}{d}\Big) \nonumber .\\ 
        & & \label{ut2}
\end{eqnarray}

	\subsection{Numerical results for the transient flow between parallel plates}
	
	We now examine the time evolution of the velocity at various distances 
	from the lower plate, for different parameter values. Figure~\ref{fig1} shows velocity profiles versus time, confirming the boundary conditions. To measure elastic effects, we define the overshoot parameter
	\[
	\delta = \frac{u_M - u_S}{u_S},
	\]
	where $u_M$ is the maximum velocity reached at a point and $u_S$ is the final steady velocity there. For a Newtonian fluid, $\delta=0$ since no overshoot occurs. Thus $\delta$ quantifies the departure from Newtonian behaviour due to elasticity, since it's produced by the release of elastic energy stored at the beginning of the flow.
	
	In panel (a) of Fig.~\ref{fig23}, the mid-gap velocity ($y=d/2$) is plotted versus time for several values of $\Gamma$, with $W$ fixed. The overshoot changes dramatically with $\Gamma$, even though $W$ remains constant. This demonstrates that $W$ alone does not determine the strength of elastic effects, in agreement with our earlier analysis.

\begin{figure}[htb!]
		\centering
		\includegraphics[width=1.0\linewidth,keepaspectratio]{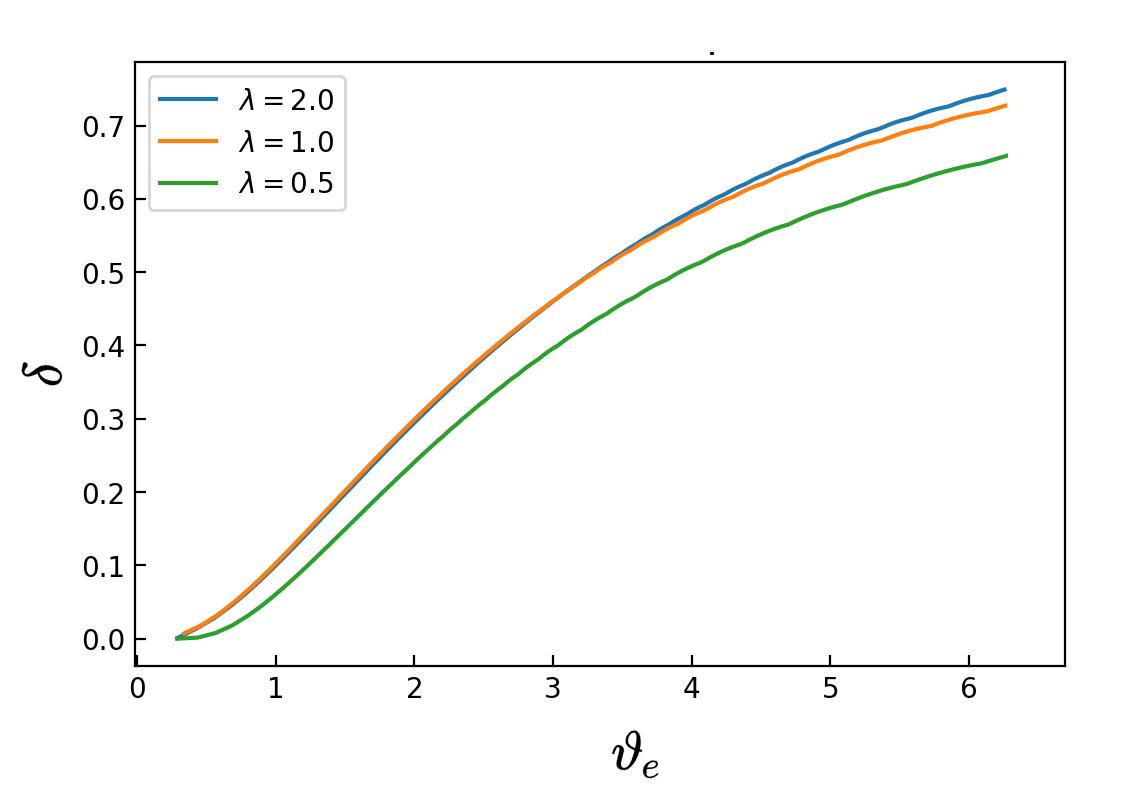}
		
		\caption{Overshoot parameter $\delta$ (see Fig.~\ref{fig1}) as a function 
		of the fluid elasticity parameter $\vartheta_e$, for different values of $\lambda$. 
		Each curve varies $G$ along it to change $\vartheta_e$. The parameter 
		$\delta$ increases nearly linearly with $\vartheta_e$.}
		\label{fig4}
	\end{figure}

    Next, we examine how $\delta$ correlates with $Wi$ and $\vartheta_e$.
    Figure~\ref{fig4} displays $\delta$ as a function of $\vartheta_e$ for various values of $\lambda$, maintaining $W$ constant along each curve (by varying $G$). At a fixed Reynolds number, the overshoot increases almost linearly with $\vartheta_e$, whereas the influence of $W$ is relatively weak; consequently, $\delta$ is essentially determined by $\vartheta_e$.
    
\begin{figure}[htb!]
		\centering
		\includegraphics[width=1.0\linewidth,keepaspectratio]{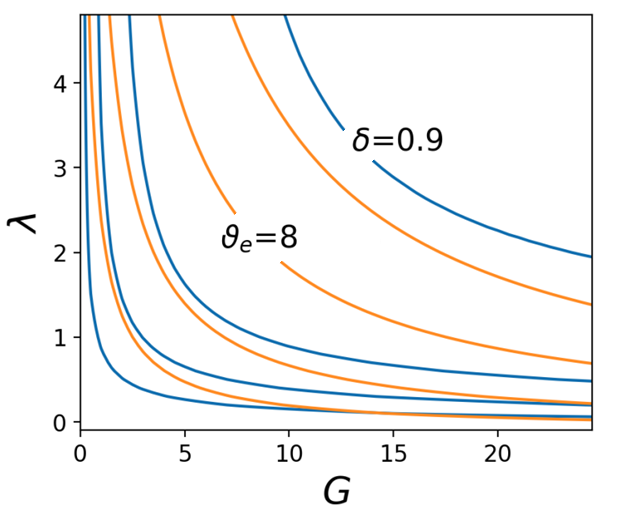}
		
		\caption{Colour map of the overshoot as a function of $\lambda$ and $G$ for $Re=1$. Blue solid lines represent curves of constant $\vartheta_e$ for values (from bottom to top) $\vartheta_e = 3, 5, 8, 11$. The orange lines correspond to level curves of $\delta$ for values (from bottom to top) $0.3, 0.5, 0.7, 0.9$. These curves are nearly parallel, indicating that the overshoot is proportional to $\vartheta_e$.}
		\label{fig2D}
	\end{figure}

    Figure~\ref{fig2D} displays the level curves of $\delta(\lambda,G)$ (dashed lines) together with the contours of constant $\vartheta_e$ (solid lines). These sets of curves are nearly parallel, confirming that the overshoot is proportional to $\vartheta_e$.

\begin{figure}[htb!]
		\centering
		\includegraphics[width=1.0\linewidth,keepaspectratio]{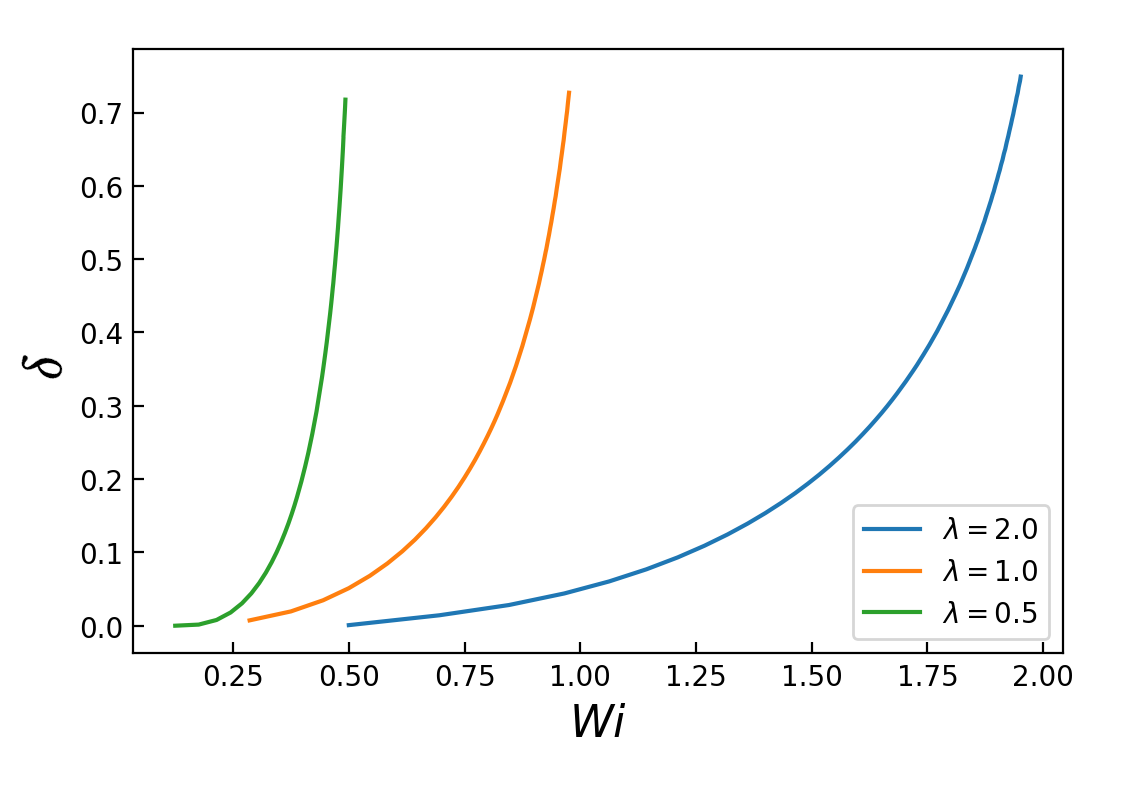}
		
		\caption{ 
			Overshoot parameter as a function of $Wi$ for different values of $\lambda$, which implies that  the curves correspond to distinct values of $W$.  }
		\label{fig5}
	\end{figure}

    Figure~\ref{fig5} presents $\delta$ as a function of $Wi$ for different values of $\lambda$ (and hence different $W$). The overshoot increases monotonically with $Wi$; however the specific relationship depends on $\lambda$.

	\subsection{Flow between concentric cylinders}
	
	To further probe the role of $\vartheta_e$, we performed numerical simulations of start-up flow of an Oldroyd-B fluid between concentric cylinders (inner radius $r_1=R$, outer radius $r_2=2R$). The fluid and cylinders are initially at rest. At $t=0$ the inner cylinder is impulsively started with an angular velocity $\Omega$. The governing Eqs.~(\ref{b1})--(\ref{ob3}) were solved using the finite-element package COMSOL~\cite{comsol}, which has been extensively validated for viscoelastic flow simulations (cf.\ Refs.~\cite{Martel, Zhu, Jensen, Obembe, Ren}). 
    Grid independence was established by employing three distinct meshes consisting of $2^{13}$, $2^{14}$, and $2^{15}$ elements, respectively. The results exhibited negligible sensitivity to mesh resolution, with maximum deviations in the azimuthal velocity remaining below 0.6$\%$.

\begin{figure}[htb!]
		\centering
		\includegraphics[width=1.0\linewidth,keepaspectratio]{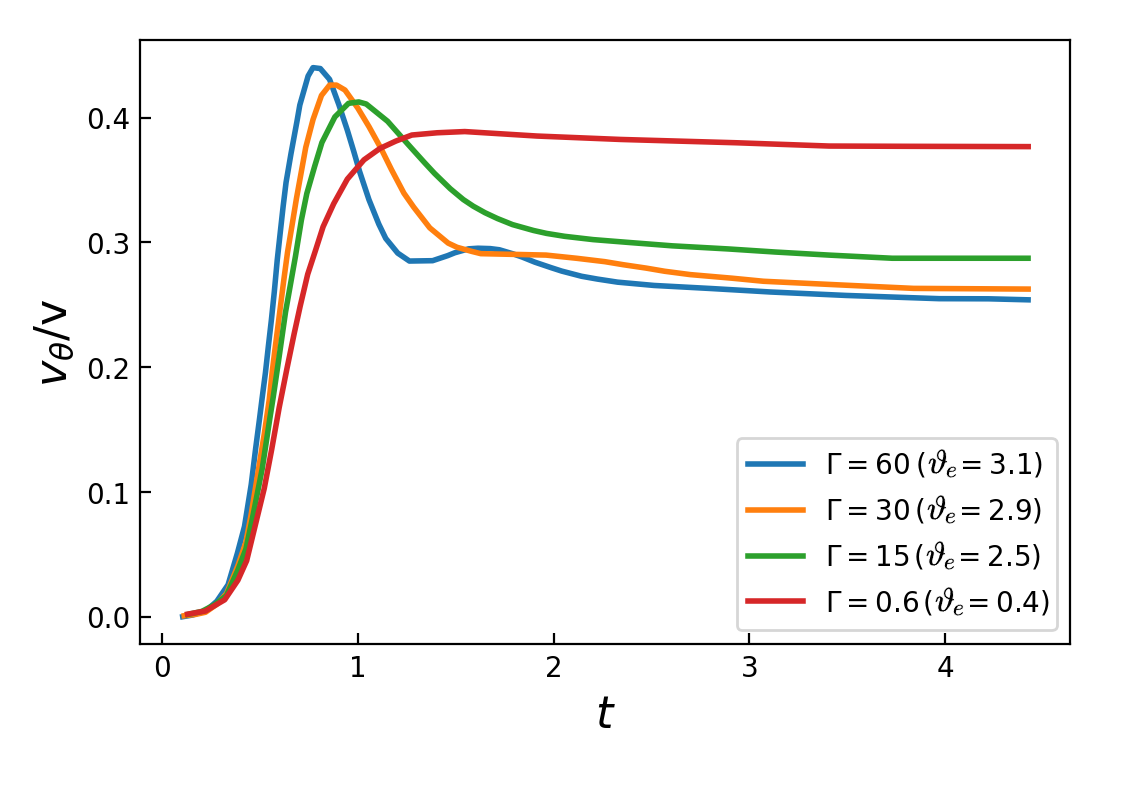}
		
		\caption{Temporal variation of the dimensionless azimuthal velocity 
		$v_\theta/U$ of the flow between cylinders for $W=0.2$ and different values of $\Gamma$ (hence different $\vartheta_e$). The labels show the corresponding values of $\vartheta_e$. It can be observed that the elastic effects range from weak ($\Gamma=0.6$) to strong ($\Gamma=60$), even though $W$ is low. In the case $\Gamma=0.6$, the response is nearly that of a Newtonian fluid.}
		\label{fig6}
	\end{figure}

	Figure~\ref{fig6} shows the time evolution of the dimensionless azimuthal velocity $v_\theta/(\Omega R)$ at $r=1.5R$, for $W=0.2$ and various $\Gamma$. The flow transitions from strong elastic oscillations at $\Gamma=60$ to much weaker oscillations at lower $\Gamma$, even though $W$ is fixed. This agrees with the parallel-plate results, indicating that $W$ alone does not control elastic strength. Again, larger $\vartheta_e$ corresponds to stronger elastic overshoot.
	
	\section{Other constitutive models}\label{SecOtherModels}
	
	Although our analysis was based on Oldroyd-B, it extends to other models. In the FENE-P model, which includes the effect of finite extensibility, one finds that as $G\to0$ the elastic term vanishes, even if $De=\lambda U/L$ is finite. Thus varying $G$ changes elastic effects dramatically at fixed $De$. Therefore, as with Oldroyd-B, the value of $De$ does not determine elastic importance in FENE-P.
	
	Similarly, in the Giesekus model (expressed with ${\bf s}=G(\sigma-I)$), one must recover Oldroyd-B in the limit $\alpha \rightarrow 0$, so that, we have $G=\mu_p/\lambda$. Substituting in Eqs.~(\ref{Giesekus21})--(\ref{Giesekus2}), and dividing by $\mathrm{G}$ we obtain:
	\begin{equation}
		{\rho \frac{D {\bf v}}{Dt}=  -\nabla p + \mu_s \nabla^2 {\bf v} + \ \! \mathrm{G} \nabla \cdot  {\bf \sigma},}
		\label{Giesekus4}
	\end{equation}
	\begin{equation}
		\lambda \stackrel{\nabla}  {\bf \sigma} + \  {\bf \sigma-I}+ \alpha   ({\bf \sigma-I})^ 2=  0 .
		\label{Giesekus5}
	\end{equation}
    The governing Eqs.~(\ref{Giesekus4})--(\ref{Giesekus5}) then show that if $G\to0$, the elastic forces disappear even for arbitrarily large $\lambda$. Thus in this model too $De$ does not by itself set the elastic level.
	
	Finally, in the linear PTT model (Eqs.~(\ref{PTTmodel})), if one expresses again the polymeric tensor as ${\bf s}=\mathrm{G} (\sigma-I)$, the governing equations of the model results:
	\begin{equation}
		\rho \frac{D {\bf v}}{Dt}=  -\nabla p + \mu_s \nabla^2 {\bf v} + \ \! \mathrm{G} \nabla \cdot  {\bf \sigma},
	\end{equation}
	\begin{equation}
		\lambda \stackrel{\nabla}  {\bf \sigma} + \ (\sigma - {\bf I}) (1-3\epsilon+\epsilon \mathrm{Tr}(\sigma)) =2\lambda {\bf E}.
	\end{equation}
	Thus, setting $G=0$ removes elastic stress regardless of $\lambda$. Hence again $De$ is not a reliable indicator of elasticity. These examples suggest that parameters involving $G$ (such as $Wi$ and $\vartheta_e$) are needed to characterise elastic effects properly.
	For instance, $Wi$ provides an order-of-magnitude estimate of the elastic stresses in the flow, whereas $\vartheta_e$ serves as a measure of the viscoelastic properties of the fluid.
	
	\section{Conclusions}\label{SecConclusions}

In this work we have shown that, contrary to what is
frequently assumed in the literature, the Deborah number
$De$ and the {\it kinematical}  Weissenberg number defined as $W = \lambda \dot{\gamma} $,
do not by themselves determine the relative importance
of elastic forces in the Oldroyd-B model and related constitutive
descriptions. This conclusion follows from an
analysis of both the non-dimensionalization of the governing
equations and the microscopic origin of the model
parameters. As discussed above, a suitable parameter
characterizing the relevance of elastic effects should vanish
in the\textit{ limit  where elastic forces vanish, i.e., $G \rightarrow 0$.}
However, this requirement is not fulfilled by $De$, which
may remain finite even when the elastic contribution becomes
negligible.

These considerations are further supported
by the analysis of two transient flow configurations,
namely flows between parallel plates and between
concentric cylinders. In these examples we examined the
influence of parameters such as $\Gamma$, $\vartheta_e$, and  the 
two different definitions of the Weissenberg number, $Wi$ and $W$, on
the resulting viscoelastic dynamics. 
Although increasing $W=\lambda\dot{\gamma}$ enhances the elastic response, elastic effects vanish as $G \rightarrow 0$ (i.e. $\Gamma\to0$) for a fixed $W$, indicating that $W$ alone is insufficient. While the Weissenberg number $W_i$ (Eq.~\ref{dwi}) effectively captures the influence of elasticity on the flow, its relationship with the observed overshoot remains dependent on $\lambda$.

The parameter $\vartheta_e$, obtained from the nondimensional analysis and the definition of $Wi$, is independent of the flow conditions and represents an intrinsic property of the fluid that quantifies its tendency to exhibit viscoelastic effects. Indeed, the analytical solution reveals an approximately linear relationship between $\vartheta_e$ and the overshoot $\delta$ (Fig.~\ref{fig4}), provided $Re$ is kept constant. Consequently, a combined approach using $W_i$ and $\vartheta_e$ proves most effective: $W_i$ provides an order-of-magnitude estimate of elastic effects in a specific flow, whereas $\vartheta_e$ constitutes a material property that characterises the fluid's elasticity. This yields a practical simplification; to analyse different flow regimes in a given configuration, one need not explore independent variations of $\lambda$ and $G$, as varying $\vartheta_e$ suffices. In all transient flow cases examined, the observed elastic effects were found to be proportional to both $Wi$ and $\vartheta_e$.

 In conclusion, when employing the Oldroyd-B model or related constitutive frameworks, parameters such as $Wi$ and $\vartheta_e$ provide a more appropriate characterization of flow-induced elastic effects and the intrinsic viscoelasticity of the fluid. As shown in this study, suitable metrics for these effects must necessarily depend on both $\lambda$ and $G$.

\end{document}